\documentclass[aps,prl,twocolumn,showpacs,superscriptaddress]{revtex4}

\newcommand{\ra}{\rangle}
\newcommand{\la}{\langle}

\usepackage{amsmath}
\usepackage{graphicx}

\def\bbbone{{\mathchoice {\rm 1\mskip-4mu l} {\rm 1\mskip-4mu l}
{\rm 1\mskip-4.5mu l} {\rm 1\mskip-5mu l}}}
\def\Tr{\mbox{Tr}}

\begin{document}

\title{Generation of high-fidelity four-photon cluster state
and quantum-domain demonstration of one-way quantum computing 
} 

 \author{Yuuki Tokunaga}
 \email[]{tokunaga.yuuki@lab.ntt.co.jp}
 \affiliation{Division of Materials Physics, Graduate school of Engineering Science, Osaka University, Toyonaka, Osaka 560-8531, Japan}
 \affiliation{NTT Information Sharing Platform Laboratories, NTT Corporation, 
3-9-11 Midori-cho, Musashino, Tokyo 180-8585, Japan}
 \affiliation{CREST Photonic Quantum Information Project, 4-1-8 Honmachi, Kawaguchi, Saitama 331-0012, Japan}

 \author{Shin Kuwashiro}
 \affiliation{Division of Materials Physics, Graduate school of Engineering Science, Osaka University, Toyonaka, Osaka 560-8531, Japan}
 \affiliation{CREST Photonic Quantum Information Project, 4-1-8 Honmachi, Kawaguchi, Saitama 331-0012, Japan}

 \author{Takashi Yamamoto}
 \affiliation{Division of Materials Physics, Graduate school of Engineering Science, Osaka University, Toyonaka, Osaka 560-8531, Japan}
 \affiliation{CREST Photonic Quantum Information Project, 4-1-8 Honmachi, Kawaguchi, Saitama 331-0012, Japan}

 \author{Masato Koashi}
 \affiliation{Division of Materials Physics, Graduate school of Engineering Science, Osaka University, Toyonaka, Osaka 560-8531, Japan}
 \affiliation{CREST Photonic Quantum Information Project, 4-1-8 Honmachi, Kawaguchi, Saitama 331-0012, Japan}

 \author{Nobuyuki Imoto}
 \affiliation{Division of Materials Physics, Graduate school of Engineering Science, Osaka University, Toyonaka, Osaka 560-8531, Japan}
 \affiliation{CREST Photonic Quantum Information Project, 4-1-8 Honmachi, Kawaguchi, Saitama 331-0012, Japan}
\date{\today}

\begin{abstract} 
We experimentally demonstrate a simple  
scheme for generating a four-photon entangled cluster state
with fidelity over 0.860 $\pm$ 0.015.
We show that the fidelity is high enough to guarantee that 
the produced state is distinguished from GHZ, $W,$ 
and Dicke types of genuine four-qubit entanglement. 
We also demonstrate basic operations of one-way quantum computing 
using the produced state 
and show that the output state fidelities surpass classical bounds,  
which indicates that the entanglement in the produced state
essentially contributes to the quantum operation.

\end{abstract}



\maketitle
 
%=========================
% Introduction
%=========================

There has been much interest in special multi-partite entangled states, 
called cluster states, because they are used as a resource for 
one-way quantum computing (QC),
in which computation proceeds 
by a sequence of single-qubit measurements with classical feedforward
\cite{106-Raussendorf2001}.
Recently, several schemes for preparing cluster states were
demonstrated \cite{592-WALTHER2005,694-Kiesel2005,703-Walther2005,704-Prevedel2007,706-Lu2007,707-Vallone2007,716-Chen2007}.
They have shown that the produced states have genuine 
multi-partite entanglement and/or given
the proof-of-principle demonstration of one-way QC.

In this work, 
we report an experimental demonstration of
a simple scheme for preparing a four-photon cluster state 
\begin{equation}
\begin{split}
|C_4 \ra =&\frac{1}{2}(|H\ra_1|H\ra_2|H\ra_3|H\ra_4 
+|H\ra_1|H\ra_2|V\ra_3|V\ra_4 \\
&+|V\ra_1|V\ra_2|H\ra_3|H \ra_4 -
|V\ra_1|V\ra_2|V\ra_3|V\ra_4). \label{C4}
\end{split}
\end{equation} 
Here, $|H \ra$ ($|V \ra$) represents the state of a photon with horizontal
(vertical) polarization.
The state fidelity of the produced state 
was over 0.860 $\pm$ 0.015. 
This guarantees that not only the produced state 
has genuine four-qubit entanglement, 
but also the state is distinguished from  
classes of genuine four-qubit entangled states
including GHZ, $W,$ and Dicke types of entangled states. 
In order to distinguish the produced state from four-qubit Dicke states,
the state fidelity should be over 0.75 \cite{697-Tokunaga2006}, 
which was not achieved in previous four-photon experiments
\cite{592-WALTHER2005,694-Kiesel2005,703-Walther2005,704-Prevedel2007}.
Using the high-fidelity cluster states, 
we also demonstrated basic operations of 
one-way quantum computing and obtained high fidelities for 
output states.
Existing demonstrations of one-way QC 
\cite{592-WALTHER2005,694-Kiesel2005,704-Prevedel2007}  
gave the output state fidelities of quantum operations 
only as numeric data. 
We further evaluate whether the high fidelities of the output states 
really come from the existence of 
entanglement of cluster states or not.
For that purpose, 
we propose a classical bound on the fidelity as a solid benchmark 
for entanglement-based quantum information processing.
Then, we show that our experimental results of the basic operations 
of one-way QC surpass the classical bounds, 
which indicates that
the entanglement of cluster states really contributes to
one-way QC.
The benchmark can be generally useful for one-way QC
and other kinds of experiments of quantum information processing.

\begin{figure}
 \includegraphics[scale=0.40]{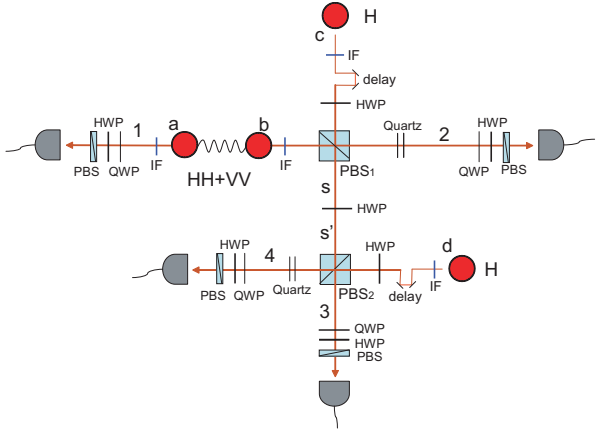}
\caption{\label{scheme}
Experimental setup for preparing $|C_4\ra$.
}
\end{figure}

Our scheme for preparing $|C_4 \ra$  
(Fig.\ \ref{scheme}) is a slight modification of the scheme
for preparing 
$|\chi\ra=\frac{1}{2}[(|HH\ra+|VV\ra)|HH\ra+(|HV\ra+|VH\ra)|VV\ra]$
in \cite{603-Tokunaga2005}, 
which is a resource for teleportation-based 
controlled-{\sc not} gate \cite{001-GC1999}.
Here, $|H \ra$ ($|V \ra$) 
represents the state of a photon with horizontal
(vertical) polarization. 
Our scheme has fewer requirements and/or a greater success probability 
compared to the schemes for existing four-photon experiments
\cite{592-WALTHER2005,694-Kiesel2005,703-Walther2005,704-Prevedel2007}.
It is constructed from four photons 
produced by parametric down-conversion (PDC), 
polarizing beam splitters (PBSs), half-wave plates (HWPs), 
and conventional photon detectors.
It does not need polarization dependent beam splitters 
\cite{694-Kiesel2005}, 
nor the subwavelength stability of the optical paths
\cite{592-WALTHER2005,703-Walther2005,704-Prevedel2007}.

%=========================
% Experimental setup
%=========================
\begin{figure}
 \includegraphics[scale=0.35]{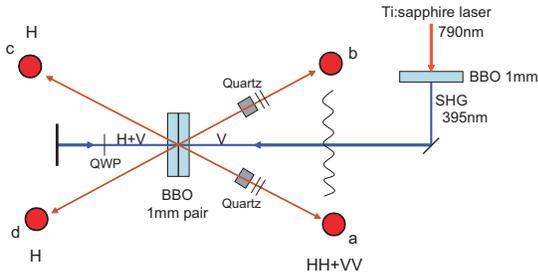}
\caption{\label{PDC}
Experimental setup for preparing two single photons and
an entangled photon pair.
}
\end{figure}
We use spontaneous parametric down-conversion for the
preparation of the entangled photon pair \cite{091-Kwiat1999}
and two single photons (Fig.\ \ref{PDC}).
Ultraviolet pulse with a central wavelength of 395 nm,
an average power of 220 mW  
from a frequency-doubled mode-locked Ti:sapphire laser
(wavelength, 790nm; pulse width, 140 fs; repetition rate, 76MHz) 
pumps a pair of 1mm-thick BBO ($\beta$-Barium Borate, type-I) 
crystal for PDC. 
The group delay is compensated by thick quartz crystals (12.8 mm) to erase
the information on the origin (the first or the second BBO) of the 
photon pairs. 
The relative phase between $H$ and $V$ polarizations
is adjusted by a pair of thin quartz crystals (0.6 mm).
The typical two-fold coincidence rate of entangled photon pairs is around
$2500$/s and the visibility is $\sim97\%$.
The temporal overlap is adjusted by moving mirrors on
motorized stages (delays in Fig.\ \ref{scheme}). 
The thin quartz crystal pairs 
in modes $2$ and $4$ are placed 
to compensate additional phase shifts.
The spectral filtering is achieved with narrow bandwidth interference
filters (IFs) with bandwidth of 2.7 nm (FWHM). 
Photon detectors (silicon avalanche photodiodes) are placed after
single-mode optical fibers to select a single spatial mode to 
ensure a high visibility.
Polarization correlations are recorded by coincidence counting among four
photon detectors for various angles of quarter-wave plates (QWPs) 
and HWPs.
The typical four-fold coincidence rate is around 100 per hour.

%=================================
% fidelity lower bound
%=================================
We obtain lower bounds on the fidelity $F$ of the produced state
using methods with fewer measurement settings 
\cite{538-Toth2005,686-Toth2005,697-Tokunaga2006} 
compared to the method for obtaining the exact fidelity.
We denote $X$, $Y$, and $Z$ for Pauli matrices $\sigma_x$,
$\sigma_y$, $\sigma_z$, respectively.
When a self-adjoint operator $B$ satisfies 
\begin{equation}
|C_4 \ra\la C_4| \ge B, \label{ineq} 
\end{equation}
we can obtain a lower bound on the fidelity 
by $F \equiv \Tr[|C_4 \ra\la C_4|\rho]\ge \la B \ra \equiv \Tr[B\rho]$ 
\cite{697-Tokunaga2006}.
The operators 
\begin{equation}
\begin{split}
B_{2}:=\frac{1}{4}(&ZZII+IZXX+ZIXX \\
&+XXZI+IIZZ+XXIZ)-\frac{1}{2}\bbbone, 
\end{split} \label{B2X}
\end{equation}
and
\begin{equation}
\begin{split}
B_4 := \frac{1}{8}(&XXZI+IZXX+ZIXX+XXIZ \\
&-YYZI-IZYY-ZIYY-YYIZ)
\end{split} \label{B4}
\end{equation}
satisfy Eq. (\ref{ineq}).
Therefore, we can obtain lower bounds on the fidelity 
by measuring expectation values
$\la B_2 \ra $ or $\la B_4 \ra $.
We need two measurement settings $XXZZ$ and $ZZXX$
for $B_2$ \cite{538-Toth2005,686-Toth2005}, and four measurement settings 
$XXZZ$, $ZZXX$, $YYZZ$, and $ZZYY$ for $B_4$ \cite{697-Tokunaga2006}.
We can obtain a higher lower bound of the fidelity using $B_4$.

\begin{figure}
 \includegraphics[scale=0.4]{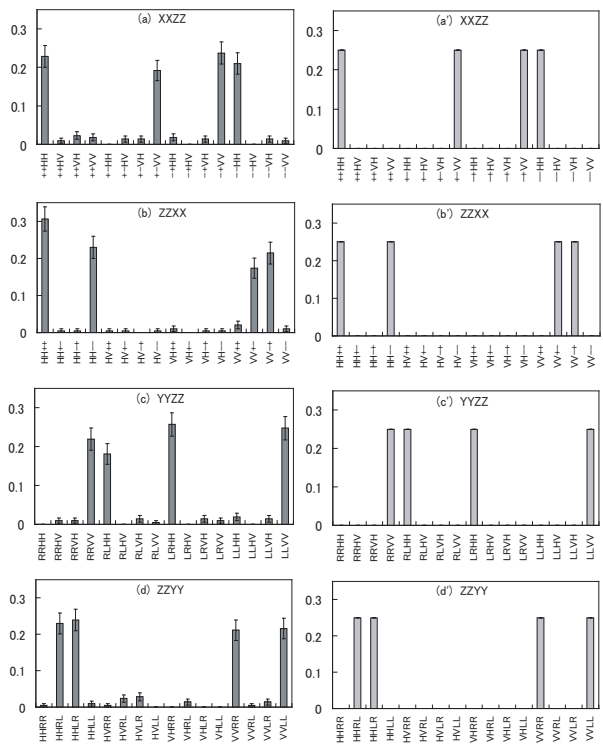}
 \caption{\label{fourfold} Four-fold coincidence probabilities 
for measurement settings
(a) $XXZZ$, (b) $ZZXX$, (c) $YYZZ$, and (d) $ZZYY$.
The ideal cases are shown in (a') - (d'). 
}
\end{figure}

%=========================
% results
%=========================
Figure \ref{fourfold}(a) - \ref{fourfold}(d) show the 16 possible fourfold coincidence 
probabilities for measurement settings $XXZZ$, $ZZXX$, $YYZZ$, and $ZZYY$, 
respectively. 
Here we denote 
$|\pm\ra \equiv \frac{1}{\sqrt{2}}(|H\ra \pm |V\ra)$, 
and $|R / L \ra \equiv \frac{1}{\sqrt{2}}(|H\ra \pm i |V\ra)$.
Figure \ref{fourfold} (a') - \ref{fourfold}(d') show the corresponding 
coincidence probabilities 
for the ideal pure four-photon cluster state $|C_4\ra$.
The error bars are determined by assuming Poissonian counting statistics. 
Deviation from the ideal case is mainly due to 
imperfection of indistinguishability of photons and
multiphoton emission events with five or more photons.
From the four-fold coincidence probabilities 
in Fig.\ \ref{fourfold}(a) and \ref{fourfold}(b), 
we can calculate $\la B_2 \ra $  to 
obtain $F \ge \Tr[B_{2}\rho]= 0.791 \pm 0.030$.
If we use all the data in Fig. \ref{fourfold}(a) - \ref{fourfold}(d) for the four settings,
we obtain a much higher bound
$F \ge \Tr[B_{4}\rho]= 0.860 \pm 0.015$.

%==============================
% witness operator, fidelity
%==============================

The observed fidelity $F>1/2$ assures that 
the produced state has genuine four-qubit entanglement. 
Moreover, the high fidelity enables us 
to discriminate it against several types of 
genuine four-qubit entangled states.
In Ref. \cite{697-Tokunaga2006}, it was shown that we can 
discriminate classes of genuine four-qubit entanglement by
extending Schmidt number witness \cite{690-Sanpera2001} to multiqubit systems.
Here we sketch the main idea of the discrimination method.
Consider the three ways of partitioning the four qubits  
$1$, $2$, $3$, and $4$ into two pairs of qubits, 
$(12)(34)$, $(13)(24)$, and $(14)(23)$. 
For simplicity, we denote them as $12$, $13$, and $14$, respectively.
A pure state of the four qubits can be regarded as a bipartite state 
for partition $1j$, and let $r_{1j}$ be its Schmidt rank.
Since the Schmidt rank never increases under local operations and
classical communication
even probabilistically, the set of the ranks $(r_{12},r_{13},r_{14})$
is a signature of the state that can be regarded 
as a (crude) measure of entanglement. The cluster state 
$|C_4\ra$ has the signature $(2,4,4)$, while 
the four-qubit GHZ state $|GHZ \ra = \frac{1}{\sqrt{2}}(|0000\ra +|1111\ra)$
and the $W$ state $|W \ra = \frac{1}{2}(|0001\ra+|0010\ra+|0100\ra+|1000\ra)$
both have $(2,2,2)$. The difference can be detected via the fidelity
as follows. It was shown \cite{697-Tokunaga2006} that 
for any state $|\eta_2\ra$ with $r_{13}\le 2$ or $r_{14}\le 2$,
its fidelity to the cluster state $|\la \eta_2| C_4\ra|^2$
is not greater than 1/2. Hence the observed fidelity of $F>1/2$
assures that the produced state is never written as 
a mixture of states with $r_{13}\le 2$ or $r_{14}\le 2$,
including the GHZ states and the $W$ states. 
Similarly, any state with $r_{13}\le 3$ or $r_{14}\le 3$
has fidelity not greater than $3/4$, and hence 
$F>3/4$
assures that the produced state is never written as 
a mixture of states with $r_{13}\le 3$ or $r_{14}\le 3$,
including a four-qubit Dicke state
$|D_4 \ra = \frac{1}{\sqrt{6}}(|0011\ra +|0101\ra + |0110\ra
+|1001\ra +|1010\ra + |1100\ra)$ having the signature 
$(3,3,3)$. 
The experimentally obtained fidelity, $F \ge 0.860 \pm 0.015$, 
thus discriminates the produced state from 
the classes of entangled states with the Schmidt rank
less than 4 in partition 13 or 14,
which include GHZ, $W,$ and Dicke types of entangled states.

%============================================
%One-way computation
%============================================
Next, we report demonstration of basic operations of  
one-way quantum computing using the produced four-photon cluster state.
What we try to demonstrate here is that the entanglement in the 
produced four-photon state really contributes to 
basic operations of one-way QC.
In one-way QC, 
the entanglement in the cluster state enables us to obtain 
the correct output states with the help of the classical feedforward 
communication. If it were not for the quantum entanglement, 
it would be impossible to achieve the correct output states 
for many kinds of gate instructions 
with the same amount of classical communication. 
This leads to a classical bound on the average fidelity 
when no entanglement exists
between input qubits (for gate instruction) and output qubits. 
In the following, we first explain the implementation of 
basic operations in one-way QC.
Then we introduce the classical bounds and show that 
the experimental results are beyond the classical bounds.

\begin{figure}
   \includegraphics[scale=0.43]{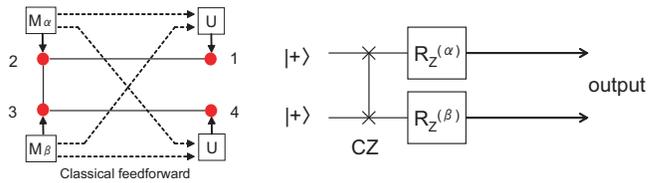}
   \caption{\label{two-qubit-fig} 
Physical implementation and the quantum circuit of a two-qubit gate.}
\end{figure}

{\it Two-qubit gates.}
We implement the quantum circuit in Fig.~\ref{two-qubit-fig} 
via one-way quantum computing model using  $|C_4 \ra$.
This implementation is basically the same as 
\cite{592-WALTHER2005,704-Prevedel2007}.
The input state is $|\psi_{\text{in}}\ra=|+ \ra|+ \ra$. 
Qubits 2 and 3 are measured in the basis
$B(\alpha)$ and $B(\beta)$, where
$B(\theta)=\{\frac{|0\ra + e^{-i\theta}|1\ra}{\sqrt{2}}, \frac{|0\ra -
e^{-i\theta}|1\ra}{\sqrt{2}}\}$.
Here we take $\{|0\ra \equiv |H\ra, |1\ra \equiv |V\ra\}$
as a standard basis.
The outcomes are 
feedforwarded and Pauli operations are applied on qubits 1 and 4
accordingly, resulting in the output state 
$|\psi_{\text{out}}\ra=(R_Z(\alpha)\otimes R_Z(\beta))\text{CZ}|\psi_{\text{in}}\ra$ on qubits 1 and 4. 
Here, $R_Z(\theta)=\exp(-i\theta\sigma_Z/2)$ 
and CZ operation is defined as 
$|j\ra|k\ra \mapsto (-1)^{jk}|j\ra|k\ra$, where $j,k \in 0,1$.
Note that the gate instructions $(\alpha, \beta)$ are given to 
qubits 2 and 3, and only two bits are communicated to 
qubits 1 and 4. 
As in the remote state preparation (RSP) protocols \cite{140-Bennett2001},
the entanglement in the cluster state enables us to obtain 
the correct output states 
with the help of the two-bit communication.
If it were not for the entanglement,
achieving the correct output states for all values of $(\alpha, \beta)$
would be impossible, leading to a bound on the average fidelity. 
In experiment, we chose 8 combinations of 
$(\alpha, \beta)$, and measured the fidelity of the output
states (the feed-forwarded Pauli operations are substituted by the 
appropriate reassignment of measurement bases for qubits 1 and 4). 
Table \ref{two-qubit} shows the results. 
\begin{table}
\caption{Output state fidelities of two-qubit gates.}
\label{two-qubit}
\begin{center}
\begin{tabular}{cccc} \hline \hline
$\alpha$ & $\beta$ & \ \ \ \ \ \ \ \ \ \ \ \ \ Output state \ \ \ \ \ \ \ \  \ \ &  Fidelity \\ \hline 
0 & 0 & $|\psi_1\ra=|H\ra|+\ra+|V\ra|-\ra$ & 0.831 $\pm$ 0.033 \\
0 & $\pi/2$ & $|\psi_2\ra=|H\ra|R\ra+|V\ra|L\ra$ & 0.847 $\pm$ 0.036\\
0 & $\pi$ & $|\psi_3\ra=|H\ra|-\ra+|V\ra|+\ra$ &  0.924 $\pm$ 0.025\\
0 & $-\pi/2$ & $|\psi_4\ra=|H\ra|L\ra+|V\ra|R\ra$ & 0.899 $\pm$ 0.028\\ 
$\pi$ & 0 & $|\psi_5\ra=|H\ra|+\ra-|V\ra|-\ra$ & 0.912 $\pm$ 0.028\\
$\pi$ & $\pi/2$ & $|\psi_6\ra=|H\ra|R\ra-|V\ra|L\ra$ &  0.913 $\pm$ 0.028\\
$\pi$ & $\pi$ & $|\psi_7\ra=|H\ra|-\ra-|V\ra|+\ra$ &  0.925 $\pm$ 0.024\\
$\pi$ & $-\pi/2$ & $|\psi_8\ra=|H\ra|L\ra-|V\ra|R\ra$ &  0.910 $\pm$ 0.027\\
\hline \hline
\end{tabular}
\end{center}
\end{table}
Let us determine the upper bound on the average fidelity 
when we do not have entanglement at all. The only 
clue about which of the 8 operations 
are chosen at qubits 
2 and 3 is the two-bit signal sent to qubits 1 and 4.
Hence a possible strategy is to divide the 8 states into 4 groups, 
e.g., (i) $|\psi_1\ra$, $|\psi_2\ra$,
(ii) $|\psi_3\ra$, $|\psi_4\ra$,
(iii) $|\psi_5\ra$, $|\psi_6\ra$, and 
(iv) $|\psi_7\ra$, $|\psi_8\ra$, and to 
send the identity of the group. Using this information, 
qubits 1 and 4 are prepared in one of the states
(i) $|\psi_1\ra + |\psi_2\ra$,
(ii) $|\psi_3\ra + |\psi_4\ra$,
(iii) $|\psi_5\ra + |\psi_6\ra$, and
(iv) $|\psi_7\ra + |\psi_8\ra$,
which were chosen such that the best average fidelity is 
achieved for each group
(normalization factors omitted).
This particular strategy gives the average fidelity of the 8 states 
$\cos ^2 (\pi/8)\approx 0.854$. Since the statistical 
 mixture of strategies does not improve the optimal fidelity,
the possible strategies are exhausted 
by all the combinations of the grouping of eight states, which 
is finite, and we have exhaustively confirmed that the above number is the 
optimal one. On the other hand, average of the eight fidelities in
Table \ref{two-qubit} gives 0.895 $\pm$ 0.010, indicating that 
our demonstration of the one-way QC achieved the fidelity that 
is only possible through the contribution of the entanglement in
the produced cluster state.

\begin{figure}
   \includegraphics[scale=0.50]{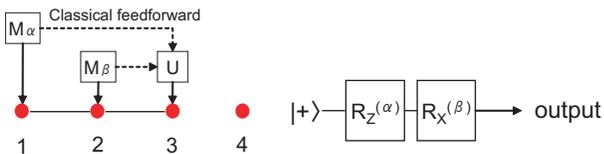}
   \caption{\label{single-fig2}
Physical implementation and the quantum circuit of a single-qubit rotation.}
\end{figure}

{\it Single-qubit rotations.} 
The quantum circuit in Fig.\ \ref{single-fig2} shows 
a simple implementation of a single-qubit rotation.
Qubit 4 is disentangled 
from the cluster state by measuring in the basis $\{|+\ra, |-\ra\}$.
The input state is $|\psi_{\text{in}}\ra=|+ \ra$. 
Qubits 1 and 2 are measured in the basis 
$B'(\alpha)$ and $B(\beta)$, respectively, where
$B'(\theta)=\{\frac{|+\ra + e^{-i\theta}|-\ra}{\sqrt{2}}, \frac{|+\ra -
e^{-i\theta}|-\ra}{\sqrt{2}}\}$.
The outcomes are 
feedforwarded and Pauli operations are applied on qubits 3 
accordingly, resulting in the output state 
$|\psi_{\text{out}}\ra =R_X(\beta)R_Z(\alpha)|+\ra$,
where $R_X(\theta)=\exp(-i\theta\sigma_X/2)$. 
The gate instructions $(\alpha, \beta)$ are given to 
qubits 1 and 2, and only two bits are communicated to 
qubit 3. 
In experiment, we chose 6 combinations of 
$(\alpha, \beta)$, and measured the fidelity of the output
states.
Table \ref{single} shows the results. 
\begin{table}
\caption{Output state fidelities of single-qubit rotations.}
\label{single}
\begin{center}
\begin{tabular}{ccccc} \hline \hline
$\alpha$ & $\beta$ & \ \ \ \ \ Output state \ \ \ &  Fidelity \\ \hline
0 & 0 &  $|+\ra$ & 0.944 $\pm$ 0.022 \\
$\pi$ & 0 &  $|-\ra$ & 0.888 $\pm$ 0.029 \\
$\pi/2$ & 0 &  $|R\ra$ & 0.928 $\pm$ 0.026 \\
$-\pi/2$ & 0 &  $|L\ra$ & 0.969 $\pm$ 0.017 \\
$\pi/2$ & $\pi/2$ & $|H\ra$ & 0.915 $\pm$ 0.029 \\
$\pi/2$ & $-\pi/2$ &  $|V\ra$ & 0.917 $\pm$ 0.027 \\ \hline \hline
\end{tabular}
\end{center}
\end{table}
As in the case of two-qubit gates, we determine the upper bound 
on the average fidelity when we do not have entanglement at all. 
An optimal strategy is to divide the 6 states into 4 groups, 
e.g., 
(i) $|H\ra$, (ii) $|V\ra$, (iii) $|+\ra$, $|R\ra$, and
(iv) $|-\ra$, $|L\ra$, and 
to send the identity of the group. Using this information, 
qubit 3 is prepared in one of the states
(i) $|H\ra$, (ii) $|V\ra$, (iii) $|+\ra+|R\ra$,
or (iv) $|-\ra+|L\ra$ 
(normalization factors omitted).
This strategy gives the average fidelity of the 6 states 
$(2/6) \times 1 + (4/6) \times \cos ^2 (\pi/8)\approx 0.902$.  
On the other hand, average of the six fidelities in
Table \ref{single} gives 0.926 $\pm$ 0.010,
indicating that the entanglement in the produced state
really contributes to the one-way QC.

%==========================================
%\Conclusions
%==========================================
We have demonstrated a high-fidelity four-photon cluster state
that is distinguished from other types of genuine four-qubit entanglement
such as GHZ, $W$, and Dicke states.
We have also shown that the results of the basic operations 
of one-way QC surpass the classical bounds,
which indicates that
the entanglement of cluster states really contributes to
one-way QC.
The model of one-way QC is unique in that the computation process is 
divided into preparation of a nonlocal static resource (a cluster state) 
and dynamic execution involving only local measurements and classical 
communication. This gives a close link to the quantum communication 
problems, and various classical bounds related to communication tasks such 
as the proposed bound here, which may be called ``classical RSP bound'', 
will be used as benchmarks toward the 
realization of quantum computing. The relation between such bounds and the 
computational power is also an interesting problem, which may give us a 
deeper insight into the role of entanglement in the quantum computation.

\begin{acknowledgments}
We thank \c{S}ahin K. \"{O}zdemir, Ryo Namiki,
Shigeki Takeuchi, Ryo Okamoto, and Tomohisa Nagata, 
for their helpful advice and discussions.
This work was supported by 21st Century COE Program by the Japan Society 
for the Promotion of Science 
and a MEXT Grant-in-Aid for Young Scientists (B) 17740265.
\end{acknowledgments}

%\bibliography{quantum}

\end{document}